\documentclass[12pt]{article}
\usepackage[utf8]{inputenc}
\usepackage{authblk}
\usepackage{amsmath}
\usepackage{amsfonts}
\usepackage{amssymb}
\usepackage{multirow}
\usepackage{comment}
\usepackage{lscape} 
\usepackage[round]{natbib}
\usepackage{url} 
\usepackage{graphicx}
\usepackage{subcaption}
\usepackage{xcolor}
\usepackage{url}

\usepackage[margin = 1in]{geometry}

\DeclareMathOperator*{\argmax}{arg\,max}
\DeclareMathOperator*{\argmin}{arg\,min}
\def\bs{\boldsymbol}
\def\bb{\mathbb}

\newtheorem{prop}{Proposition}

\title{A Compound Decision Approach to Covariance Matrix Estimation}

\author{Huiqin Xin and Sihai Dave Zhao}
\affil{Department of Statistics, University of Illinois at Urbana-Champaign}
\begin{document}


\maketitle

\label{firstpage}
\begin{abstract}
  Covariance matrix estimation is a fundamental statistical task in many applications, but the sample covariance matrix is sub-optimal when the sample size is comparable to or less than the number of features. Such high-dimensional settings are common in modern genomics, where covariance matrix estimation is frequently employed as a method for inferring gene networks. To achieve estimation accuracy in these settings, existing methods typically either assume that the population covariance matrix has some particular structure, for example sparsity, or apply shrinkage to better estimate the population eigenvalues. In this paper, we study a new approach to estimating high-dimensional covariance matrices. We first frame covariance matrix estimation as a compound decision problem. This motivates defining a class of decision rules and using a nonparametric empirical Bayes $g$-modeling approach to estimate the optimal rule in the class. Simulation results and gene network inference in an RNA-seq experiment in mouse show that our approach is comparable to or can outperform a number of state-of-the-art proposals.
\end{abstract}

\noindent
{\small
Keywords: Compound decision theory; $g$-modeling; nonparametric maximum likelihood; separable decision rule.
  }

\section{Introduction}
\label{introduction}

Covariance matrix estimation is a fundamental statistical problem that plays an essential role in various applications. However, in modern problems the number of features can be of the same order as or exceed the sample size. This high-dimensional setting is especially common in genomics, where covariance matrices are used to model gene networks but the number of genes can be much larger than the number of biological replicates \citep{schafer2005shrinkage, markowetz2007inferring}. For example, in Section \ref{gene analysis} we study brain region-specific gene networks in mouse using bulk RNA-sequencing data. A common approach to estimating gene expression networks is to use the standard sample covariance or corrlation matrices \citep{langfelder2008wgcna, zhang2005general}. However, these estimators behave poorly in the high-dimensional regime, and there are roughly two types of approaches to address this issue.

Structured methods make assumptions about the form of the population covariance matrix. One popular class of methods assumes that the matrix is sparse, or has many zero entries. The most common strategy in this class is to threshold the entries of the sample covariance \citep{rothman2009generalized, cai2011adaptive}, but penalized likelihood methods \citep{xue2012positive} have also been used. A second class of methods assume the data arise from a factor model \citep{fan2008high}, so that the covariance matrix has low intrinsic dimension. Other common structured methods assume that the covariance matrix is banded \citep{li2017estimation} or Toeplitz \citep{liu2017covariance}.

In contrast, unstructured methods do not make any assumptions about the population covariance matrix, yet still have lower estimation error than the sample covariance matrix. Many of these methods can be interpreted as rotationally invariant estimators, which shrink the sample eigenvalues in a data-adaptive fashion \citep{bun2016rotational, stein1975estimation, stein1986lectures}. One example is the linear shrinkage approach of \citet{ledoit2004well}, which shrinks the eigenvalues toward their global mean. Some more recently developed nonlinear shrinkage methods have been shown to be asymptotically optimal in this class \citep{ledoit2012nonlinear, ledoit2019quadratic, lam2016nonparametric}.

Though these optimal estimators perform extremely well, \citet{bun2016rotational} showed that the eigenvectors of any rotationally invariant estimate must be the same as those of the sample covariance matrix. This can be problematic because sample eigenvectors are not consistent when the dimension and the sample size increase at the same rate \citep{mestre2008asymptotic}. This suggests that unstructured but non-rotationally invariant covariance matrix estimators may be worth exploring, as these can modify both the sample eigenvectors as well as the eigenvalues. The difficulty is to identify a reasonable class of estimators to pursue.


In this paper we define and explore one such class of estimators. This approach is based on interpreting the covariance matrix estimation problem as a compound decision problem \citep{robbins1951asymptotically}. Section \ref{sec:compound} shows how this framing motivates vectorizing the covariance matrix and solving the resulting vector estimation problem using nonparametric empirical Bayes procedures studied in the compound decision literature. The vector estimate can then be reassembled into matrix form and projected into the space of positive-definite matrices to give a final estimate. Surprisingly, though this vectorized approach essentially ignores the matrix structure, it can sometimes outperform a number of state-of-the-art proposals in simulations and a real data analysis. Our compound decision approach is closely related to the correlation matrix estimator of \citet{dey2018corshrink}. We discuss in detail the connections with their work in Section \ref{sec:proposed}.

The article is organized as follows. In Section \ref{sec:compound} we briefly review compound decision theory, and in Section \ref{sec:method} we introduce our proposed estimator. In Section \ref{numerical results} we illustrate its performance in simulations and Section \ref{gene analysis} contains an analysis of the gene coexpression network inference problem. Section \ref{sec:discussion} concludes with a discussion, and proofs can be found in the Supporting Information.

\section{\label{sec:compound}Covariance matrix estimation as a compound decision problem}
\subsection{\label{sec:background}Problem formulation}

Given $n$ observations $\bs{X}_1,\ldots,\bs{X}_n$ independently generated from a $p$-dimensional $\mathcal{N}(\bs{0},\bs{\Sigma})$, our goal is to find an estimator $\bs{\delta}(\bs{X})$ of $\bs{\Sigma}$. A common measure of estimation performance is the scaled squared Frobenius risk
\begin{equation}
\label{frobenius risk}
R(\bs{\Sigma}, \bs{\delta}) = \frac{1}{p^2} \sum_{j,k=1}^p \mathbb{E}[\{\delta_{jk}(\bs{X})-\sigma_{jk}\}^2],
\end{equation}
where $\sigma_{jk}$ is the $jk$th entry of $\bs{\Sigma}$ and $\delta_{jk}(\bs{X})$ is its corresponding estimate and $\sigma_{jj} = \sigma^2_j$ is the variance of the $j$th feature.

Minimizing \eqref{frobenius risk} can be viewed as a compound decision problem. These involve the simultaneous estimation of multiple parameters $\bs{\theta} = (\theta_1, \ldots, \theta_n)^\top$ given data $\bs{Y} = (Y_1, \ldots, Y_n)^\top$, with $Y_i\sim P_{\theta_i}$ \citep{robbins1951asymptotically}. Specifically, the goal is to develop a decision rule $\bs{\delta}(\bs{Y}) = (\delta_1(\bs{Y}),\ldots,\delta_n(\bs{Y}))$ that minimizes the compound risk $ n^{-1} \sum_{i=1}^n \bb{E} L(\theta_i,\delta_i(\bs{Y}))$, where $L$ is a loss function measuring the accuracy of $\delta_i(\bs{Y})$ as an estimate of $\theta_i$. A classical example is the classical Gaussian sequence problem, where $Y_i \sim N(\theta_i, 1)$ independently and $L(t, d) = (t - d)^2$ is the squared error loss \citep{johnstone2017gaussian}.

Covariance matrix estimation under the Frobenius risk \eqref{frobenius risk} is thus a compound decision problem where the goal is to simultaneously estimate the components of
\begin{equation}
  \label{eq:sigma}
  \bs{\sigma} = (\sigma_{11}, \ldots, \sigma_{1p}, \ldots, \sigma_{p1}, \ldots,\sigma_{pp})^\top
\end{equation}
under average squared error loss.  One important difference between estimating a covariance matrix versus a vector is that the former should have additional structure, and in particular should be at least positive semidefinite. Notably, however, this structure is not incentivized by the Frobenius risk \eqref{frobenius risk}. As a result, there can exist estimators of $\bs{\Sigma}$ that achieve low values of \eqref{frobenius risk} but which are not positive-definite. To resolve this issue, we can project any estimate into the space of positive-definite matrices; see Section \ref{posdef}.

The value of this formulation is that it allows us to apply ideas from the compound decision literature to covariance matrix estimation.  A key property of compound decision problems is that while information about a parameter $\theta_i$ seems like it should offer no help for estimating any other $\theta_j$, in fact borrowing information across all $\theta_i$ is superior to considering each parameter separately. An example of this phenomenon is the James-Stein shrinkage estimator \citep{james1961estimation}
, and a long line of subsequent work has led to much more sophisticated and accurate procedures \citep{brown2009nonparametric, jiang2009general, johnstone2017gaussian, lindley1962discussion, fourdrinier2018shrinkage}. In Section \ref{sec:method} we apply these ideas to covariance matrix estimation.

\subsection{\label{sec:connections}Reinterpretation of existing methods}

Treating covairance matrix estimation as a vector estimation problem may seem unintuitive, but many existing matrix estimation methods can also be reinterpreted as carrying out vector estimation. The simplest example is the sample covariance matrix $\bs{S} = n^{-1} \bs{X}^{T}\bs{X}$, which takes this form because the $\bs{X}$ are assumed to have mean zero. As another example, \citet{cai2011adaptive} studied sparse high-dimensional covariance matrices and explicitly appealed to the vector perspective. Their adaptive thresholding method applies a version of the soft thresholding method of \citet{donoho1995adapting}, originally developed to estimate the mean of a Gaussian sequence, to each entry of the sample covariance matrix.

As a further illustration of the potential of our new formulation, we can show that the linear shrinkage covariance matrix estimator of \citet{ledoit2004well} can be essentially recovered as the solution to a compound decision problem. A standard approach in this literature is to first define a class of decision rules for estimating each component and then identify the member of that class that minimizes an estimate of the compound risk 
\citep{fourdrinier2018shrinkage, stigler19901988}. Following this recipe, we consider the class of linear decision rules where $\sigma_{jk}$ will be estimated by
\begin{equation}
  \label{linear class}
  \delta_{jk}(\bs{X}) = \beta_S s_{jk} + \beta_I u_{jk}
\end{equation}
for some scalar parameters $\beta_S$ and $\beta_I$, where $s_{jk}$ is the $jk$th entry of $\bs{S}$ and $u_{jk}$ is the $jk$th entry of the $p \times p$ identity matrix $\bs{I}$. Ideally we would like to estimate the optimal $\beta_S$ and $\beta_I$ by choosing the values that minimize the Frobenius risk \eqref{frobenius risk}. Though we cannot directly calculate \eqref{frobenius risk} because it involves the unknown $\sigma_{jk}$, the following proposition shows that we can construct a good estimate of this risk for every fixed value of $\beta_S$ and $\beta_I$.

\begin{prop}
  \label{prop:Rhat}
  Define $\hat{\Delta}_{jk}^2 = \sum_{i=1}^{n}(X_{ij}X_{ik}-s_{jk})^2 / n^2$ and
  \begin{equation}
    \label{eq:Rhat}
    \hat{R}_L(\beta_S,\beta_I) = \frac{1}{p^2} \sum_{j,k=1}^{p}[(2\beta_S-1) \hat{\Delta}_{jk}^2 + \{(1- \beta_S) s_{jk} - \beta_I u_{jk}\}^2 ].
  \end{equation}
  If $p^{-2} \bb{E} \Vert \bs{S} - \bs{\Sigma} \Vert_F^2$ is bounded as $n \rightarrow \infty$, then
  \[
    \lim_{n \rightarrow 0} \{ \bb{E} \hat{R}_L(\beta_S, \beta_I) - R(\bs{\Sigma}, \bs{\delta}) \} = 0
  \]
  for the class of decision rules $\bs{\delta}$ defined in \eqref{linear class}, where $\Vert \bs{A} \Vert_F$ denotes the Frobenius norm $\bs{A}$.
\end{prop}

The condition on $\bb{E} \Vert \bs{S} - \bs{\Sigma} \Vert_F^2$ requires that the variances of the entries of $\bs{S}$ do not grow too quickly as $n$ grows. It is also required by \citet{ledoit2004well} and is implied by their Lemma 3.1. Proposition \ref{prop:Rhat} shows that $\hat{R}_L(\beta_S, \beta_I)$ \eqref{eq:Rhat} is an asymptotically unbiased estimate of the true risk \eqref{frobenius risk} for linear decision rules of the form \eqref{linear class}. It is then reasonable to estimate the best estimator in this class by minimizing $\hat{R}_L(\beta_S, \beta_I)$ over $\beta_S$ and $\beta_I$. A slight modification of this procedure turns out to recover the \citet{ledoit2004well} estimator
\begin{equation}
\label{eq:lw}
\hat{\bs{\Sigma}}_{\text{LW}} = \left(1 - \frac{b_n^2}{d_n^2}\right) \bs{S} + \frac{b_n^2}{d_n^2}\hat{\mu} \bs{I},
\end{equation} 
where $\hat{\mu} = \text{tr}(\bs{S})/p$, $d_n^2=\Vert\bs{S}-\hat{\mu} \bs{I} \Vert_F^2$, $b_n^2 = \min(d_n^2, \sum_{i=1}^n \Vert \bs{X}_i \bs{X}_i^\top - \bs{S} \Vert_F^2 / n^2)$.

\begin{prop}
\label{prop:linear}
Let $\hat{\beta}_S$ and $\hat{\beta}_I$ denote the minimizers of $\hat{R}_L(\beta_S, \beta_I)$ and define the linear decision rule
\begin{equation}
  \label{eq:opt_linear}
  \hat{\delta}_{jk}(\bs{X}) = \max(\hat{\beta}_S, 0) s_{jk} + \min(\hat{\beta}_I, \hat{\mu}) u_{jk}.
\end{equation}
Then $\hat{\delta}_{jk}(\bs{X}) = \hat{\sigma}_{\text{LW}jk}$, where $\hat{\sigma}_{\text{LW}jk}$ is the $jk$th entry of $\bs{\Sigma}_{\text{LW}}$ \eqref{eq:lw}.
\end{prop}


\section{\label{sec:method}Methods}
\subsection{\label{sec:class}New class of estimators}

Proposition \ref{prop:linear} of Section \ref{sec:connections} suggests that applying ideas from the compound decision literature to estimating the vector $\bs{\sigma}$ \eqref{eq:sigma}  can be a fruitful strategy for covariance matrix estimation. However, Proposition \ref{prop:linear} considers only a linear decision rule, whereas more recent work in compound decision theory has focused on the much larger class of so-called separable rules \citep{brown2009nonparametric, jiang2009general, zhang2003compound}. 

We generalize these ideas to propose a new class of covariance matrix estimators. For rules $\bs{\delta}(\bs{X}) = (\delta_{11}(\bs{X}),\ldots,\delta_{pp}(\bs{X}))$ that estimate $(\sigma_{11}, \ldots, \sigma_{pp})^\top$, we define the class of separable estimators to be
\begin{equation}
  \label{separable}
  \mathcal{S} = \{\bs{\delta} :
  \quad
  \delta_{kj} = \delta_{jk} = t_{od}(\bs{X}_{\cdot j}, \bs{X}_{\cdot k}),\, 1\leq k<j\leq p;
  \quad 
   \delta_{jj} = t_d(\bs{X}_{\cdot j}),\,j=1,\ldots,p\},
\end{equation}
where $\bs{X}_{\cdot j} = (X_{1j}, \ldots, X_{nj})^\top$ is the vector of observed values of the $j$th feature. In other words, rules in $\mathcal{S}$ estimate diagonal entries of $\bs{\Sigma}$ using a function $t_d$ and off-diagonal entries using $t_{od}$, where $t_{od}$ and $t_d$ do not depend on the indices $j$ and $k$. Furthermore, we enforce rules in $\mathcal{S}$ to give symmetric estimates of the off-diagonal entries.


This class of separable rules is reasonable to consider because it includes several common covariance estimators, including the sample covariance, the class of adaptive thresholding estimators for sparse covariance matrices studied by \citet{cai2011adaptive}, and the class of linear estimators \eqref{linear class} used by \citet{ledoit2004well}, which can be expressed as
\begin{align*}
t_{od}(\bs{X}_{\cdot j}, \bs{X}_{\cdot k}) &= \beta_S \bs{X}_{\cdot j}^\top \bs{X}_{\cdot k} / n, \, 1\leq k<j\leq p,\\
t_d(\bs{X}_{\cdot j})&=\beta_S \bs{X}_{\cdot j}^\top \bs{X}_{\cdot j} / n+\beta_I, \, j=1,\ldots,p.
\end{align*}
Furthermore, whereas these three existing estimators are ultimately all functions of the $s_{jk}$, the class $\mathcal{S}$ allows for much more general rules that can be any function of the $\bs{X}_{\cdot j}$ and $\bs{X}_{\cdot k}$. Finally, the class $\mathcal{S}$ constitutes a fundamentally different approach to covariance matrix estimation compared to the existing methods described in Section \ref{introduction}. Estimators in $\mathcal{S}$ do not assume that $\bs{\Sigma}$ has any particular structure and also are not necessarily rotationally invariant.


\subsection{\label{sec:proposed}Proposed estimator}

We propose to search for the optimal estimator $\bs{\delta}^\star$ within $\mathcal{S}$ \eqref{separable}. We can first provide a closed-form expression for $\bs{\delta}^\star$. Let $f_2(\cdot \mid a, b, \gamma)$ be the density of a bivariate normal centered at zero with standard deviations $a$ and $b$ and correlation $\gamma$, $f_1(\cdot \mid a)$ be the density of a univariate normal with mean zero and standard deviation $a$, and
\begin{equation}
  \label{eq:G}
  \begin{aligned}
    G_{od}(a, b, \gamma)
    =\,&
    \frac{1}{p (p - 1)} \sum_{1 \leq j, k \leq p, j \ne k} I(\sigma_j \leq a, \sigma_k \leq b, r_{jk} \leq \gamma),\\
    G_d(a)
    =\,&
    \int dG_{od}(a,b,\gamma) db d\gamma = \frac{1}{p} \sum_{j=1}^p I(\sigma_j \leq a),
  \end{aligned}
\end{equation}
where $\sigma_j$ and $\sigma_k$ are the true standard deviations of the $j$th and $k$th covariates and $r_{jk} = \sigma_{jk} / (\sigma_j \sigma_k)$ is their true correlation.

\begin{prop}\label{prop:bayes risk}
  The optimal separable estimator
  \begin{equation}
    \label{eq:opt}
    \bs{\delta}^\star
    =
    \argmin_{\bs{\delta} \in \mathcal{S}} R(\bs{\Sigma}, \bs{\delta})
    =
    \left\{
      \begin{array}{ll}
        t_{od}^\star(\bs{X}_{\cdot j}, \bs{X}_{\cdot k}), & 1 \leq k < j \leq p,\\
        t_d^\star(\bs{X}_{\cdot j}), & j = 1, \ldots, p,
      \end{array}
    \right.
  \end{equation}
  obeys
  \begin{equation}
    \label{eq:tstar}
    \begin{aligned}
      t_{od}^\star(\bs{X}_{\cdot j}, \bs{X}_{\cdot k})
      =\,&
      \frac{\int a b \gamma f_{2n}(\bs{X}_{\cdot j}, \bs{X}_{\cdot k} \mid a, b, \gamma) dG_{od}(a, b, \gamma)}{\int f_{2n}(\bs{X}_{\cdot j}, \bs{X}_{\cdot k} \mid a, b, \gamma) dG_{od}(a, b, \gamma)},\\
      t_d^\star(\bs{X}_{\cdot j})
      =\,&
      \frac{\int a^2 f_{1n}(\bs{X}_{\cdot j} \mid a) dG_d(a)}{\int f_{1n}(\bs{X}_{\cdot j} \mid a) dG_d(a)}
    \end{aligned}
  \end{equation}
  where
  \begin{equation}
    \label{eq:f}
    \begin{aligned}
      f_{2n}(\bs{X}_{\cdot j}, \bs{X}_{\cdot k} \mid a, b, \gamma)
      =
      \prod_{i = 1}^n f_2(X_{ij}, X_{ik} \mid a, b, \gamma),
      \quad
      f_{1n}(\bs{X}_{\cdot j} \mid a)
      =
      \prod_{i = 1}^n f_1(X_{ij} \mid a).
    \end{aligned}
  \end{equation}
\end{prop}

Proposition \ref{prop:bayes risk} illustrates the key property of compound decision problems: that borrowing information across indices can help estimate a single parameter. In this case, the optimal separable rule for estimating the $jk$th entry of $\bs{\Sigma}$ borrows from all of the $\sigma_j$ and $r_{jk}$. This rule also has an interesting Bayesian interpretation, as $t^\star_{od}$ and $t^\star_d$ are formally equivalent to posterior expectations when the $(\sigma_j, \sigma_k, r_{jk})$ are independently and identically distributed according to the prior $G_{od}$. However, the result in Proposition \ref{prop:bayes risk} holds even when the parameters are not random. Proposition \ref{prop:bayes risk} is a direct extension of the fundamental theorem of compound decisions \citep{robbins1951asymptotically, zhang2003compound}.

The optimal $\bs{\delta}^\star$ \eqref{eq:opt} is an oracle estimator and cannot be calculated in practice. A common approach in the compound decision literature is to use empirical Bayes ideas to estimate the oracle optimal separable decision rule \citep{robbins1955empirical, zhang2003compound, brown2009nonparametric, jiang2009general, efron2014two, efron2019bayes}. We propose to estimate $\bs{\delta}^\star$ using what \citet{efron2014two} calls $g$-modeling, where we proceed as if the $(\sigma_j, \sigma_k, r_{jk})$ were truly random and then estimate the prior $G_{od}$ from the data. Specifically, under the working assumption that the $(\bs{X}_{\cdot j},\bs{X}_{\cdot k})$ are independent for different $(j, k)$, we propose to estimate $G_{od}$ and $G_d$ \eqref{eq:G} using nonparametric maximum likelihood \citep{kiefer1956consistency}:
\begin{equation}
  \label{Gnd hat}
  \hat{G}_{od} = \argmax_{G_{od} \in \mathcal{G}_{od}}
  \prod_{1\leq k < j \leq p} \int f_{2n}(\bs{X}_{\cdot j},\bs{X}_{\cdot k} \mid a, b, \gamma) dG_{od}(a, b, \gamma)
  \prod_{j = 1}^p \int f_{1n} (\bs{X}_{\cdot j} \mid a) dG_d(a),
\end{equation}
where $\mathcal{G}_{od}$ is the family of all distributions supported on $\mathbb{R}_+ \times \mathbb{R}_+ \times [-1, 1]$ and $G_d$ is determined by $G_{od}$ as indicated in \eqref{eq:G}. Of course, the $(\bs{X}_{\cdot j}, \bs{X}_{\cdot k})$ are not independent across different $(j, k)$ unless the true covariance $\bs{\Sigma}$ is diagonal, so $\hat{G}_{od}$ does not maximize a likelihood but rather a pairwise composite likelihood \citep{varin2011overview}. Using $\hat{G}_{od}$ and $\hat{G}_d$, we propose to estimate the vectorized $\bs{\Sigma}$ using
\begin{equation}
  \label{proposed}
  \hat{\bs{\delta}}(\bs{X})
  =
  (
  \hat{t}_d(\bs{X}_{\cdot 1}), \ldots, \hat{t}_{od}(\bs{X}_{\cdot 1}, \bs{X}_{\cdot p}), \ldots, \hat{t}_{od}(\bs{X}_{\cdot p}, \bs{X}_{\cdot 1}), \ldots, \hat{t}_d(\bs{X}_{\cdot p})
  )
\end{equation}
where $\hat{t}_{od}$ and $\hat{t}_d$ are obtained by plugging $\hat{G}_{od}$ and $\hat{G}_d$ \eqref{Gnd hat} into the expressions for $t_{od}^\star$ and $t_d^\star$ from \eqref{eq:tstar}.

Our proposed procedure is similar to the method of \citet{dey2018corshrink}, who introduce an empirical Bayes approach to estimating a correlation matrix. They first apply Fisher's $Z$-transform to each off-diagonal entry of the sample correlation matrix and model the means of the transformed entries as independent random variables arising independently from some prior distribution. They then estimate the prior from the data, calculate the posterior expectations of the means, transform them back into a correlation matrix, and select the closest positive-definite matrix.

The main difference between our approaches is that \citet{dey2018corshrink} do not shrink the variances while we do, as our $\hat{t}_d$ estimates $\sigma_j$ using its supposed posterior expectation assuming a prior distribution of $\hat{G}_d$. Furthermore, our Proposition \ref{prop:bayes risk} shows that our shrinkage approach can be interpreted as viewing the triples $(\sigma_j, \sigma_k, r_{jk})$ as if they were drawn from the trivariate prior $G_{od}$, which allows for possible dependencies between $r_{jk}$ and the $\sigma_j$. For example, model 2 in our simulations in Section \ref{numerical results} considers a covariance matrix where entries with larger standard deviations also tend to have larger correlations. In contrast to the method of \citet{dey2018corshrink}, our approach can learn this structure if it exists and take advantage of it for more accurate estimation.

\subsection{\label{implementation}Implementation}

Calculating the estimated prior $\hat{G}_{od}$ \eqref{Gnd hat} is difficult, as it is an infinite-dimensional optimization problem over the class of all probability distributions $\mathcal{G}_{od}$ supported on $\mathbb{R}_+ \times \mathbb{R}_+ \times [-1, 1]$. \citet{lindsay1983geometry} showed that the solution is atomic and is supported on at most $p(p+1)/2$ points. The EM algorithm has traditionally been used to estimate the locations of the support points and the masses at those points \citep{laird1978nonparametric}, but this is a difficult nonconvex optimization problem.

Instead, we maximize the pairwise composite likelihood over a fixed grid of support points, similar to recent $g$-modeling procedures for standard compound decision problems; this restores convexity \citep{jiang2009general, koenker2014convex, feng2018approximate}. Specifically, we assume that $G_{od}$ is supported on $D$ fixed support points $(a_\tau, b_\tau, \gamma_\tau)$, $\tau=1,\ldots, D$.  Since $G_{od}$ is symmetric in its first two coordinates, we construct the support points so that $D$ is even and for $\tau > D / 2$, $(a_\tau, b_\tau, \gamma_\tau) = (b_{\tau - D / 2}, a_{\tau - D / 2}, \gamma_{\tau - D / 2})$. We also constrain $\hat{G}_{od}$ such that the estimated mass at the point $(a_\tau, b_\tau, \gamma_\tau)$ is equal to the mass at $(b_\tau, a_\tau, \gamma_\tau)$. Finally, we use the EM algorithm to estimate the masses $\hat{w}_\tau$. Letting $\hat{w}_\tau^{(k)}$ denote the estimate of the $\tau$th mass point as the $k$th iteration, it can be shown that for $\tau = 1, \ldots, D / 2$, the update formula is
\begin{align*}
  \label{eq:iteration}
      &
      \hat{w}_{\tau}^{(k)} = \hat{w}_{\tau+D/2}^{(k)}\\
  =\,&
       \frac{2}{p(p+1)} \left[\sum_{j=1}^p \frac{\hat{w}_{\tau}^{(k-1)} \{f_{1n}(\bs{X}_j \mid a_{\tau})+f_{1n}(\bs{X}_j \mid a_{\tau+D/2})\}}{\sum_{l=1}^D \hat{w}_{l}^{(k-1)} f_{1n}(\bs{X}_j \mid a_{l})} \,+\right.\\
      &
          \left. \sum_{1 \leq k < j \leq p} \frac{\hat{w}_{\tau}^{(k-1)} \{f_{2n}(\bs{X}_j,\bs{X}_k \mid a_\tau, b_\tau, \gamma_\tau)+f_{2n}(\bs{X}_j,\bs{X}_k \mid a_{\tau+D/2}, b_{\tau+D/2}, \gamma_{\tau+D/2})\}}{\sum_{l=1}^{D}\hat{w}_l^{(k-1)} f_{2n}(\bs{X}_j,\bs{X}_k\mid a_l, b_l, \gamma_l)}\right].
\end{align*}
In practice, we center each $\bs{X}_{\cdot j}$ and instead of using the density functions $f_{1n}$ and $f_{2n}$ \eqref{eq:f}, we use the densities of their sufficient statistics: the sample variance $s_j^2$, in the case of $f_{1n}$, and the $2 \times 2$ matrices with diagonal entries $s_j^2$ and $s_k^2$ and off-diagonal entries equal to the sample covariances $s_{jk}$, in the case of $f_{2n}$.

The support points need to be carefully chosen. Ideally, the points $(a_\tau, b_\tau, \gamma_\tau)$ would densely cover the parameter space. However, using a grid of $d$ points in each dimension would require a total of $D = d^3$ points, which would incur a huge computational cost for even moderate $d$. Alternatively, we could use the so-called exemplar method \citep{saha2020nonparametric}, which sets the support points to equal the observed sample versions of the $(\sigma_j, \sigma_k, r_{jk})$. This would reduce the size of the support set, but the computation complexity would still grow like $O(p^2)$.

Instead, we use a clustering-based exemplar algorithm to further improve computational efficiency. Let $s_j$, $s_k$, and $\gamma_{jk}$ be the sample standard deviations and correlation between the $j$th and $k$th covariates. We first apply $K$-means clustering to identify $K$ cluster centroids using the $p (p-1) / 2$ lower triangular off-diagonal sample points $(s_j,s_k,\gamma_{jk})$. We then swap the first two coordinates of each point to construct another set of $K$ centroids and use these $2K$ centroids as the support points for $\hat{G}_{od}$. Section \ref{numerical results} evaluates the impact of changing $K$.

In our implementations of the EM algorithm in Section \ref{numerical results} we stopped iterating when the relative difference between two successive log-likelihoods was less than $10^{-4}$, up to a maximum of 200 iterations. Instead of EM, we attempted to maximize the composite likelihood \eqref{Gnd hat} using the interior point solver MOSEK, as advocated by \citet{koenker2014convex}, but we ran into computational difficulties because for even moderate sample sizes $n$, the values of the densities $f_{1n}$ and $f_{2n}$ at certain support points could be extremely small. These small values caused problems for the R wrapper for MOSEK. Our implementation of the EM algorithm was more robust to these small density values.

\subsection{\label{posdef}Positive definiteness correction}

Our proposed estimator \eqref{proposed} is not guaranteed to be positive-definite. To correct this, we reshape our vector estimator back into a matrix and then identify the closest positive-definite matrix. \citet{higham1988computing} and \citet{huang2017calibration} showed that the projection of a $p \times p$ symmetric matrix $\bs{B}$ onto the space of positive semi-definite matrices is
\[
P_0(\bs{B})
=
\argmin_{\bs{A}\geq 0} \Vert\bs{A}-\bs{B}\Vert_F
=
\bs{Q}\text{diag}\{\max(\lambda_1,0),\max(\lambda_2,0),\ldots,\max(\lambda_p,0)\}\bs{Q}^\top,
\]
where $\bs{Q}$ is the matrix of eigenvectors of $\bs{B}$, and $\lambda_1,\ldots,\lambda_p$ are its eigenvalues. 

To guarantee positive-definiteness, we follow \citet{huang2017calibration} and replace non-positive eigenvalues with a chosen positive value $c$ smaller than the least positive eigenvalue $\lambda_{\min}^+$, so that the corrected estimate is
\begin{equation}
  \label{near posdef}
  P_c(\bs{B})
  =
  \bs{Q}\text{diag}\{\max(\lambda_1,c),\max(\lambda_2,c),\ldots,\max(\lambda_p,c)\}\bs{Q}^\top.
\end{equation}
\citet{huang2017calibration} suggest $c_{\alpha}=10^{-\alpha}\lambda_{\min}^+$, where the parameter $\alpha$ is chosen to minimize $\Vert B - P_{c_{\alpha}}(B) \Vert + \alpha$ over a uniform partition of $\{\alpha_1,\ldots,\alpha_K\}$ of $[0,\alpha_K]$. In our implementations we used $K=20$ and $\alpha_K=10$.

\section{\label{numerical results}Numerical Results}

\subsection{\label{models}Covariance matrix models}

We considered six models for the population covariance matrix to explore the behavior of our proposed method in simulations. In each model, we generated certain parameters randomly but then fixed them across all replications of our simulations. For readability, below we give each model a short name that we will refer to in the following figures and captions.

Model 1 (sparse): we generated a sparse $\bs{\Sigma}$ where most features were not correlated. We let half of the standard deviations equal 1 and the other half equal 1.5. We set the correlation matrix following Model 2 of \citet{cai2011adaptive}, a block-diagonal matrix where the $jk$th entry of the first $p/2 \times p/2$ block was $\max(1- \vert j - k \vert / 10, 0)$ and the second $p/2 \times p/2$ block was an identity matrix.

Model 2 (hypercorrelated): we generated a $\bs{\Sigma}$ that we call ``hypercorrelated'' because its parameters are in a sense dependent on each other, as we let larger standard deviations $\sigma_j$ and $\sigma_k$ correspond to larger $r_{jk}$. Specifically, we let the first $p / 2$ standard deviations equal 1, the last $p / 2$ equal 2, and the correlation matrix have four $p/2 \times p/2$ blocks. The diagonal blocks were compound symmetric matrices with correlation parameters 0.8 and 0.2 and the off-diagonal blocks had entries equal to 0.4. As discussed in Section \ref{sec:proposed}, we expect that our proposed approach will be able to take advantage of the dependencies between the $\sigma_j$ and the $r_{jk}$, unlike the empirical Bayes approach of \citet{dey2018corshrink} that does not shrink standard deviations.

Models 3 and 4 (dense-0.7 and dense-0.9): we generated a $\bs{\Sigma}$ where the correlation between all features were 0.7 and 0.9, in contrast to our sparse Model 1. Otherwise, we generated the standard deviations as in Model 1.

Model 5 (orthogonal): instead of generating $\bs{\Sigma}$ by specifying its correlation matrix, we randomly generated a $\bs{U}$ from the Haar measure on the space of all orthonormal matrices and a vector of eigenvalues $\bs{l}$ independently generated from $\mathcal{U}(1, 4)$. We then let $\bs{\Sigma} = \bs{U}^T\text{diag}(\bs{l})\bs{U}$. This simulation setting was used in \citet{lam2016nonparametric} and \citet{ledoit2019quadratic}.

Model 6 (spiked): this setting was the same as model 5 except that $\bs{l}$ was a vector of eigenvalues where the first $3$ entries were 4,3,2 and the remaining $p - 3$ entries equaled 1, so that $\bs{\Sigma}$ was a spiked covariance matrix.

\subsection{\label{compared}Methods compared}

We compared several high-dimensional covariance matrix estimation procedures. Several require tuning parameters, which we fixed based on the original papers' recommendations: (a) Our proposed estimator \eqref{proposed} and its positive-definiteness-corrected version described in Section \ref{posdef}. In the following figures and captions we refer to these as MSG (Matrix Shrinkage via $G$-modeling) and MSGCor, respectively; (b) Adap: the adaptive thresholding method of \citet{cai2011adaptive} for sparse covariance matrices; (c) Linear: the linear shrinkage estimator of \citet{ledoit2004well} given in \eqref{eq:lw}; (d) QIS: the Quadratic-Inverse Shrinkage estimator of \citet{ledoit2019quadratic}, which performs linear shrinkage on the sample eigenvalues in inverse eigenvalue space; (e) NERCOME: the Nonparametric Eigenvalue-Regularized COvariance Matrix Estimator of \citet{lam2016nonparametric}, which randomly splits the samples into two groups to estimate eigenvectors and eigenvalues separately; (f) CorShrink: the empirical Bayes correlation matrix estimation method of \citet{dey2018corshrink}, where we estimated $\bs{\Sigma}$ by using CorShrink to estimate the correlation matrix and then scaling its entries using the sample standard deviations; and (g) Sample: the sample covariance matrix.

In addition to the above data-driven estimators, we also implemented two oracle estimators, which cannot be implemented in practice as they require the unknown $\bs{\Sigma}$: (a) OracNonlin: the optimal rotation-invariant covariance estimator, derived in \citet{ledoit2019quadratic}, of $\bs{U}^T\text{diag}(\bs{l})\bs{U}$, where $\bs{U}=(\bs{u}_1\ldots\bs{u}_p)$ is the sample eigenvector matrix and $\bs{l} = (d_1, \ldots, d_p)$ is composed of oracle eigenvalues $d_i = \bs{u}_i^T\bs{\Sigma} \bs{u}_i$; and (b) OracMSG: we implemented our proposed estimator \eqref{proposed} except that the support points are generated by clustering the true parameters $(\sigma_j,\sigma_k,r_{jk})$ instead of their sample versions. We performed clustering even though the true parameters are all known in order to reduce the number of support points to ease the computational burden.

\subsection{\label{optimalK}Clustering-based exemplar algorithm}

We first explored the consequences of choosing different numbers of cluster centroids in our clustering-based exemplar algorithm described in Section \ref{implementation}. For each model, we generated $n=100$ samples from a $p$-variate $\mathcal{N}(\bs{0}, \bs{\Sigma})$, where $p = 30, 100,$ or $200$. Although $\bb{E}\bs{X}=0$ in our setting, we assume it is unknown and use $\bs{S}=\frac{1}{n-1}(\bs{X}-\bs{\overline{X}})^\top(\bs{X}-\bs{\overline{X}})$. We generated $200$ replicates and reported median errors under the loss function
\begin{equation}
  \label{eq:sim_loss}
  \frac{1}{p} \left\{ \sum_{j,k = 1}^p (\hat{\sigma}_{jk} - \sigma_{jk})^2 \right\}^{1/2},
\end{equation}
where $\hat{\bs{\Sigma}}$ is the estimated matrix with entries $\hat{\sigma}_{jk}$ and $\bs{\Sigma}$ is the true matrix with entries $\sigma_{jk}$. We then fit our proposed method, after positive-definitness correction, for different $K$, where we let $K = rp$ for $r=2,1,0.5,0.25$.

\begin{figure}
\begin{center}
\centerline{\includegraphics[width=0.85\textwidth]{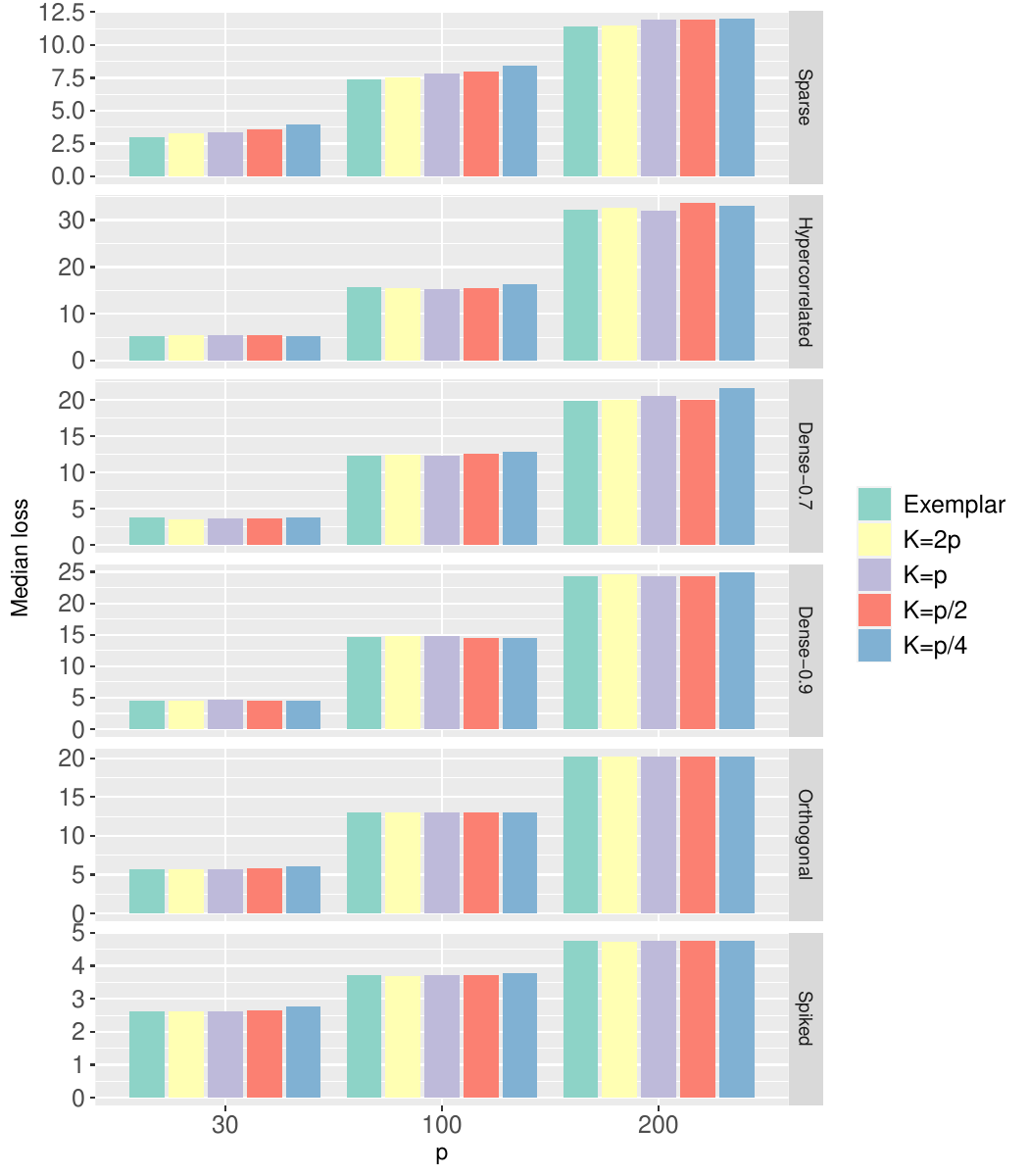}}
\end{center}
\caption{Median Frobenius norm errors over 200 replications for our proposed MSGCor. Exact numerical results and interquartile ranges are provided in the Supporting Information. Sparse: Model 1; Hypercorrelated: Model 2; Dense-0.7: Model 3; Dense-0.9: Model 4; Orthogonal: Model 5; Spiked: Model 6. This figure appears in color in the electronic version of this article, and any mention of color refers to that version.}
\label{fig:sim1_frobenius}
\end{figure}

\begin{table}
\begin{center}
\caption{\label{tab:sim1_time} Average running time of MSGCor under Model 1 for different $K$.}
\begin{tabular}{rrrr}
\hline
            & p=30 & p=100 & p=200 \\
\hline
Full example method   & 0.1357          & 6.0338        & 91.8942       \\
$K=2p$	      & 0.0539 	     & 1.1883	   & 10.7687         \\
$K=p$            & 0.0357         & 0.7852         & 7.0807         \\
$K=p/2$         & 0.0222        & 0.5231         & 4.6736         \\
$K=p/4$      & 0.0158         &0.3694          & 3.2749         \\
\hline
\end{tabular}
\end{center}
\end{table}

Figure \ref{fig:sim1_frobenius} shows that across all models, different $K$ give similar estimation accuracies compared to using all sample points in the exemplar algorithm. Table \ref{tab:sim1_time} shows that they can be significantly faster; we only provide results from Model 1 because the times did not vary much across different models. Letting $K = p$ seemed to provide a good balance between accuracy and speed, so we implement our proposed method with $K = p$ in this paper.

\subsection{\label{sec:accuracies}Estimation accuracies}

\begin{figure}
\begin{center}
\centerline{  \includegraphics[width=0.85\textwidth]{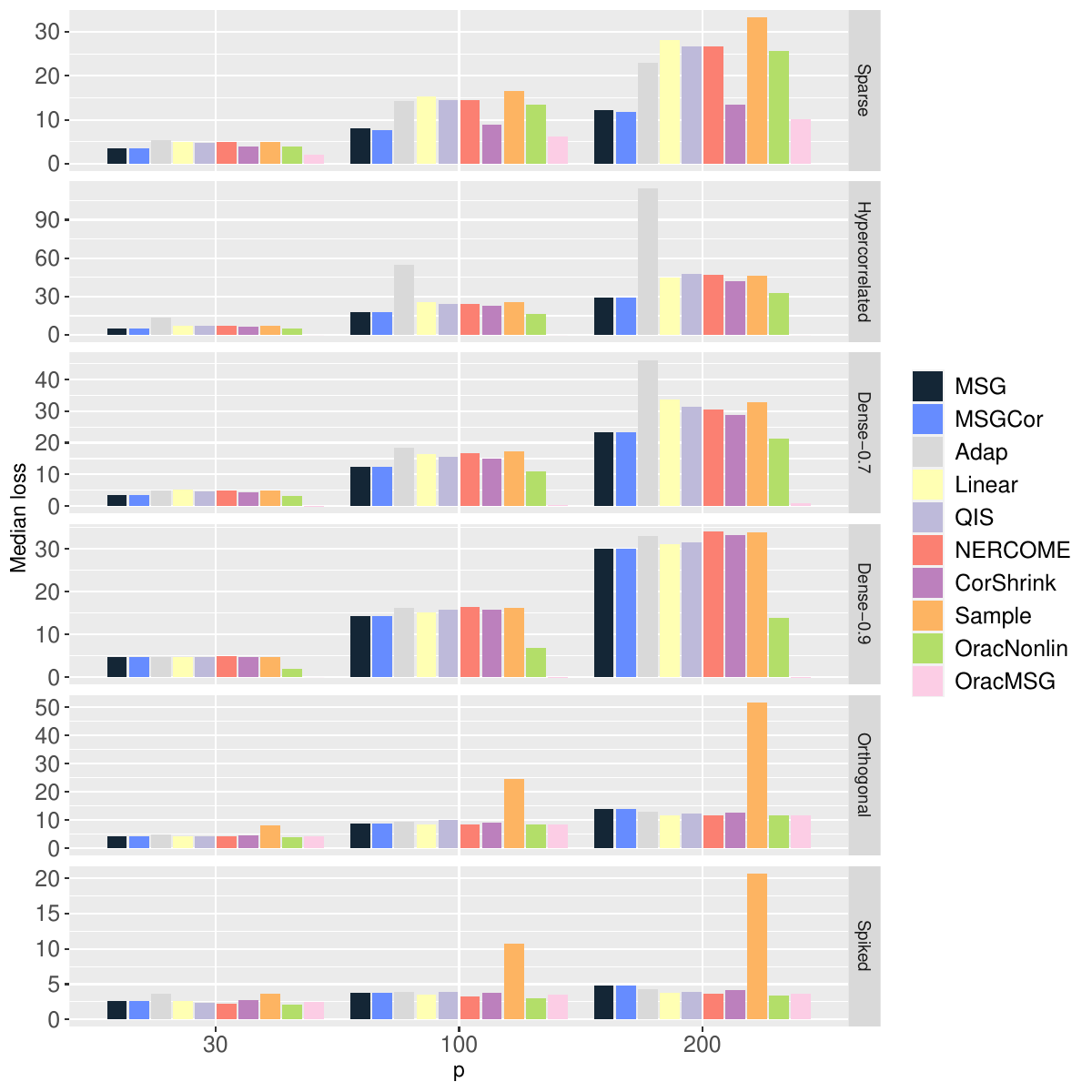}}
\end{center}
\caption{Median Frobenius norm errors over 200 replications. Exact numerical results and interquartile ranges are provided in the Supporting Information. MSG and MSGCor were implemented with $K = p$. Sparse: Model 1; Hypercorrelated: Model 2; Dense-0.7: Model 3; Dense-0.9: Model 4; Orthogonal: Model 5; Spiked: Model 6. This figure appears in color in the electronic version of this article, and any mention of color refers to that version.}
\label{fig:sim2_frobenius}
\end{figure}

We next compared the estimation accuracies of the methods in Section \ref{compared} under loss \eqref{eq:sim_loss}, following the same simulation scheme as in Section \ref{optimalK}. The results are visualized in Figure \ref{fig:sim2_frobenius}. Our proposed methods could outperform existing methods, especially in Models 1 through 4. This pattern also held when we ran simulations with $p = 1000$ in the Supporting Information. The corrected version consistently outperformed the uncorrected version. Our methods could outperform the CorShrink method of \citet{dey2018corshrink} in many cases, likely because it was able to accurately shrink the standard deviations.

MSG and MSGCor were especially good in Model 2, where the standard deviation and correlations were related, because it was able to capture this dependence in its estimate of the prior $G_{od}$ \eqref{eq:G} and leverage it to provide much more accurate estimates. It was substantially better than CorShrink in this setting, further illustrating the advantage of shrinking the standard deviations. In the Supporting Information we also compare MSG and MSGCor to CorShrink for estimating correlation instead of covariance matrices. We find that in that case, CorShrink performs slightly better than MSG and MSGCor except in Model 2, likely because the additional flexibility that our method trades off lower bias for higher variance.

However, in Models 5 and 6 MSG and MSGCor were worse than the best of the other methods, though they were still comparable. In the Supporting Information we found this pattern was exacerbated when $p = 1000$. This observation is consistent with the behavior of the oracle estimator OracMSG, which was dramatically better than the oracle rotationally-invariant estimator OracNonlin in Models 1 through 4 but was comparable to it in Models 5 and 6. We conjecture that this may be because the population eigenvectors may be closer to the sample eigenvectors in Models 5 and 6, while our class of separable estimators is best-suited for estimating covariance matrices whose sample and population eigenvalues differ greatly. We leave this for future work.


\section{Data analysis}
\label{gene analysis}

Covariance matrix estimation is often used to reconstruct gene networks \citep{markowetz2007inferring}. We used our proposed methods and the other covariance matrix estimators described in Section \ref{compared} for gene network estimation in a study of the transcriptomic response to social aggression in various brain regions in mouse \citep{saul2017transcriptional}. As part of the control condition, mice were exposed to a paper cup and bulk RNA-sequencing data was collected from the amygdala, frontal cortex, and hypothalamus at 30, 60, and 90 minutes after exposure. Data were collected from five mice at each time point, but two of the 30-minute hypothalamus samples were sequenced using a different library preparation than the other samples and so were left out of this analysis. The data are available from the NCBI Gene Expression Omnibus under accession number GSE80345.

Our goal was to estimate brain region-specific gene networks in these mice, assuming that the networks were the same across time points. Understanding how these networks differ between brain regions may provide insight into the different functions of the different regions. We built coexpression networks of the top 200 genes that were most differentially expressed between the regions. To identify these genes, we first considered only genes that were expressed in each of our samples and had at least one count per million mapped reads in at least 3 samples. We then performed a Kruskal-Wallis test and retained the 200 genes with the lowest p-values.

We then converted the gene expression values to log-counts per million mapped reads and applied the methods from Section \ref{compared} to estimate the covariance matrix of these 200 genes in each brain region. When implementing our $K$-means clustering-based exemplar algorithm, we let $K = 400$. We also only implemented MSGCor, since it was the best performer in our simulations.

\begin{table}
\begin{center}
\caption{\label{tab:tab1}Median gene expression covariance matrix estimation errors (25\% and 75\% quantiles in parentheses). Bold text highlights the smallest median errors in each column.}
\begin{tabular}{rll}
  \hline
  Brain region & Amygdala & Frontal cortex \\
  \hline
  MSGCor & \textbf{2.26 (1.98, 2.57)} & \textbf{2.3 (2.15, 2.55)} \\
  Adap & 2.6 (2.18, 2.97) & 2.44 (2.18, 2.71) \\
  Linear & 2.3 (2.05, 2.55) & 2.35 (2.19, 2.53)\\
  QIS & 2.48 (2.07, 2.85) & 2.42 (2.18, 2.70)\\
  NERCOME & 2.37 (2.14, 2.62) & \textbf{2.3 (2.14, 2.53)}\\
  CorShrink & \textbf{2.26 (2, 2.56)} & 2.36 (2.2, 2.56)\\
  Sample & 2.61 (2.33, 2.85) & 2.79 (2.64, 2.95)\\
  \hline
\end{tabular}
\end{center}
\end{table}

To measure the accuracy of the estimators, we randomly split the samples into a training and testing set. For the sake of sample size, we have ignored the timepoints at which the gene expressions were measured, following \citet{saul2017transcriptional}. We estimated the covariance matrix on the training set and measured accuracy using the Frobenius norm of the difference between the estimated matrix and the sample covariance matrix calculated from the test set. We set the training sample size equal to 10, which left five samples in the test sets for the amygdala and frontal cortex and only three samples in the test set for hypothalamus. We therefore dropped hypothalamus from the analysis because we felt that three samples was unsuitable for calculating a surrogate of the true covariance matrix. We repeated this process 100 times. Table \eqref{tab:tab1} reports the median errors and interquartile ranges across the replications. Many of the different methods had comparable performance, and the different brain regions favored different methods. Nevertheless, our proposed MSGCor had among the best performances in both regions.

\begin{figure}
  \begin{center}
    \includegraphics[width=0.45\textheight]{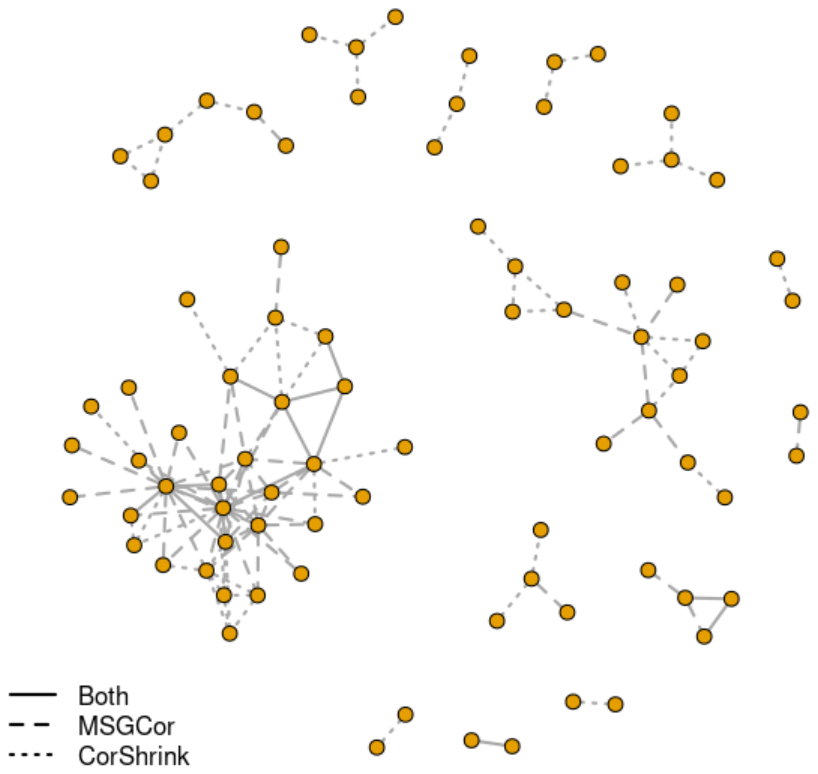}
  \end{center}
\caption{Amygdala gene networks recovered by the top performers in Table \ref{tab:tab1}. Line types indicate the method(s) that recovered each each edge. This figure appears in color in the electronic version of this article, and any mention of color refers to that version.}
\label{network_A}
\end{figure}

\begin{figure}
  \begin{center}
    \includegraphics[width=0.45\textheight]{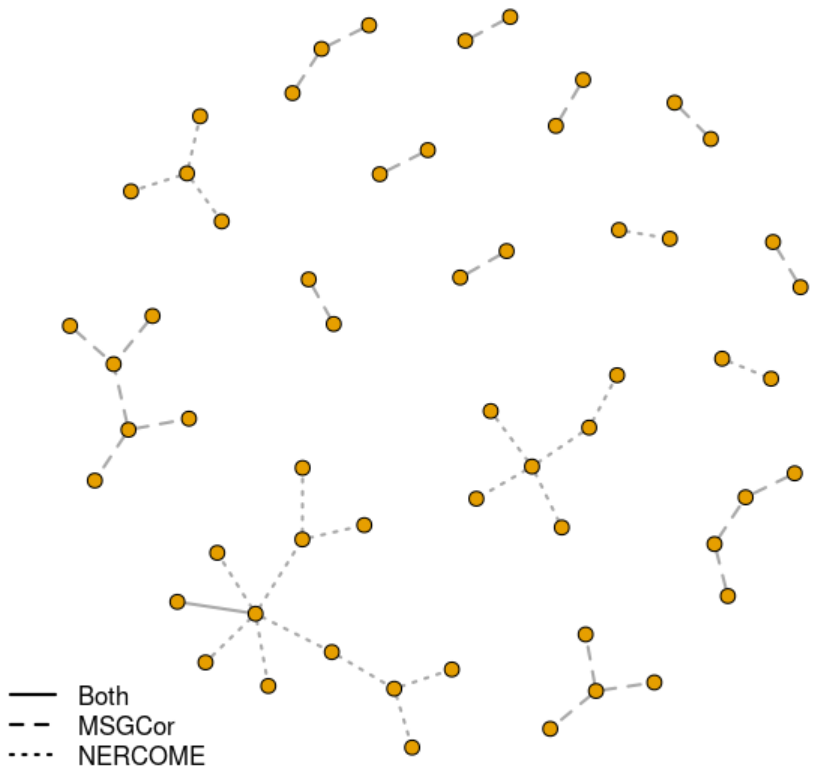}
  \end{center}
\caption{Frontal cortex gene networks recovered by the top performers in Table \ref{tab:tab1}. Line types indicate the method(s) that recovered each each edge. This figure appears in color in the electronic version of this article, and any mention of color refers to that version.}
\label{network_FC}
\end{figure}

In addition to comparing the numerical accuracies, we also compared the gene networks produced by applying the various approaches to all samples. To avoid completely connected graphs, we sparsified the matrix estimates by thresholding the smaller entries of each matrix to zero. Since the adaptive thresholding method of \citet{cai2011adaptive} naturally produced a sparse estimated matrix, we thresholded the other matrix estimates to match the sparsity level of the \citet{cai2011adaptive} estimate. Figures \ref{network_A} and \ref{network_FC} illustrate the estimated amygdala and frontal cortex covariance matrix estimates from the best performers from Table \ref{tab:tab1} in network form, where each node represents a gene and each edge represents a non-zero covariance between the genes it connects. The figures show that our MSGCor approach gave qualitatively different gene networks. This suggests a useful strategy for estimating gene networks in practice: high-confidence edges can be identified as those edges recovered by multiple covariance matrix estimators.

\section{\label{sec:discussion}Discussion}
\label{discussion}

The class of separable covariance matrix estimators \eqref{separable} that we explored in this paper appears to be promising. This is somewhat surprising because our approach vectorizes the matrix and therefore cannot take matrix structure, such as positive-definiteness, into account. This suggests that a vectorized approach combined with a positive-definiteness constraint may have improved performance. The resulting estimator would necessarily not be separable, because the estimate of the $jk$th entry would depend on more than just the $j$th and $k$th observed features, so the $g$-modeling estimation strategy is insufficient.

Providing theoretical guarantees for our estimator is difficult. In the standard mean vector estimation problem with $Y_i \sim N(\theta_i, 1)$, \citet{jiang2009general} showed that an empirical Bayes estimator based on a nonparametric maximum likelihood estimate of the prior on the $\theta_i$ can indeed asymptotically achieve the same risk as the oracle optimal separable estimator. However, this was in a simple model with a univariate prior distribution. \citet{saha2020nonparametric} extended these results to multivariate $\bs{Y}_i \sim N(\bs{\theta}_i, \bs{\Sigma}_i)$ with a multivariate prior on the $\bs{\theta}_i$, but assumed that the $\bs{Y}_i$ were independent. In contrast, our covariance matrix estimator is built from arbitrarily dependent $(\bs{X}_{\cdot j},\bs{X}_{\cdot k})$. These impose significant theoretical difficulties that will require substantial work to address; we leave this for future research.

Finally, we have so far assumed that our data are multivariate normal, though simulations in the Supporting Information suggest that our procedures can still provide relatively accurate estimates when the normality assumption is violated. To extend our procedure to non-normal data belonging to a parametric family, we can simply modify the density function $f_1$ and $f_2$ in the nonparametric maximum compositive likelihood problem \eqref{Gnd hat} and in our proposed estimator \eqref{proposed}. If $f_1$ and $f_2$ are unknown or difficult to specify, alternative procedures may be necessary to approximate the optimal separable rule.

\section*{Acknowledgments}
We thank Dr. Roger Koenker and three anonymous referees for their very valuable comments.\vspace*{-8pt}

\section*{Data Availability Statement}
The data are available from the National Center for Biotechnology Information Gene Expression Omnibus at \url{https://www.ncbi.nlm.nih.gov/geo/} under accession number GSE80345.

\bibliographystyle{abbrvnat} 
\bibliography{biblio}
\label{lastpage}
\end{document}